\documentstyle[12pt]{article}
\input epsf
\newcommand{\be}{\begin{eqnarray}}
\newcommand{\ee}{\end{eqnarray}}

  \def\beq{\begin{equation}}
  \def\eeq{\end{equation}}
  \def\beqr{\begin{eqnarray}}
  \def\eeqr{\end{eqnarray}}
  
 \def\Tr{\mbox{Tr}}

 \def\Dirac#1{#1\hskip-6pt/}

\begin{document}
\rightline{RUB-TP2-12/00}
\vspace{1cm}
\begin{center}
{\bf\Large DVCS amplitude in the parton model} \\
\vspace{0.4cm}
{\bf M.~Penttinen$^a$, M.V.~Polyakov$^{a,b}$,
A.G.~Shuvaev$^{b}$,
M.~Strikman$^c$}\\
\vspace{0.4cm}
{\it
$^a$Institut f\"ur Theoretische Physik II, Ruhr--Universit\"at Bochum,\\
 D--44780 Bochum, Germany\\
$^b$Petersburg Nuclear Physics Institute, 188350 Gatchina, Russia\\
$^c$ Department of Physics, Pennsylvania State University,
University Park, PA 16802, USA}
\end{center}
\vspace{.2cm}
\begin{abstract}
\noindent
We compute amplitude of deeply virtual Compton scattering in the parton
model. We found that the amplitude up to the accuracy $O(1/Q)$
depends on new skewed parton distributions (SPD's). These additional
contributions make the DVCS amplitude explicitly transverse.
\end{abstract}
\section*{\normalsize {\bf Introduction}}
Hard exclusive processes, such as
deeply virtual Compton scattering (DVCS), owing to the
QCD factorization theorems \cite{CFS,JiOsborn,Radyushkin,CollinsFreund}
allow to probe so-called skewed parton distributions
\cite{CFS,Radyushkin,Bartels,Dittes,Ji}.

In this note we shall study DVCS amplitude
in the parton model, our prime interest will be the question what kind
of skewed parton distributions enter DVCS amplitude
up to the order $O(1/Q)$.
Our calculations follow closely ideas of Ref.~\cite{Anikin}.
We compute the DVCS amplitude in the parton model,
where the scattering of virtual photon occurs on
a single parton on the mass-shell. In this approach the scattering
amplitude is explicitly gauge invariant (transverse).
We show that in the parton model in the case of pion target
we reproduce results of Ref.~\cite{Anikin}. Also we give
an expression for the  DVCS amplitude for an arbitrary target.

We shall see that even in the simplest parton model the handbag
contribution to the DVCS amplitude depend not only on the
matrix elements of the light-ray operators with indices
projected on the light cone direction but also on the operators
projected on the  transverse direction.  The corresponding additional
contributions make  the  DVCS amplitude explicitly transverse.  The
additional terms in the DVCS amplitude are proportional either to the
transverse component of the momentum transfer or
the transverse component of the target polarization, as it follows from
the recent operator analysis of Ref.~\cite{BR}.

Let us emphasize that calculations in the parton model we perform
here correspond to the QCD calculation of the handbag diagram in
time ordered perturbation theory in the
infinite momentum frame. Therefore our calculations
are model independent despite the  use of the
approximation of the parton model.
The advantage of such formalism is that at each step of
the calculations the amplitude is transverse (e.m. gauge invariant).
Of course, the calculations in the parton model do not allow us
to determine the contributions of the operators of the type
$\bar \psi G \psi$ to the amplitude.
In order to determine contributions of such
operators one has to perform the QCD operator analysis.

\section*{\normalsize {\bf DVCS amplitude}}

Here we give an expression for
the handbag contribution to the DVCS amplitude computed in the parton
model. The corresponding diagrams are shown in Fig.~1.
\begin{figure}[tbp]
\epsfxsize=10cm
\epsfysize=8cm
\centerline{\epsffile{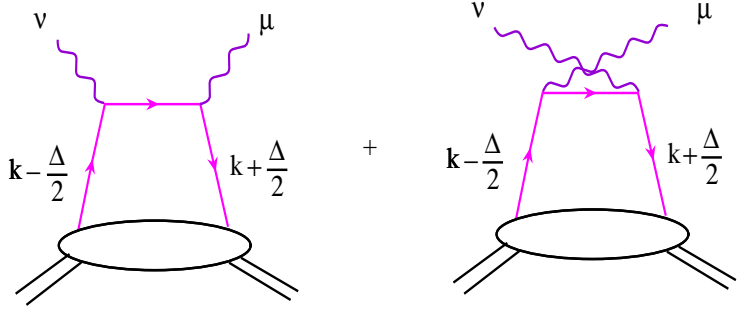}}
\caption{Handbag contribution to DVCS amplitude}
\label{iaa}
\end{figure}
It is convenient to introduce the light-cone decomposition of the momenta:
\be
\nonumber
p^\mu&=&(1+\xi)\widetilde n^\mu+
(1-\xi )\frac{\bar M^2}{2} n^\mu -\frac 12 \Delta_\perp^\mu\; \\
\nonumber
p^{\prime \mu}&=&(1-\xi )\widetilde n^\mu+
(1+\xi )\frac{\bar M^2}{2} n^\mu +\frac 12 \Delta_\perp^\mu\; \\
\nonumber
q^{\mu}&=&-2 \xi \widetilde n^\mu+
\frac{ Q^2}{4\xi} n^\mu \;\\
\nonumber
\Delta^\mu&=&(p'-p)^\mu\\
\bar M^2&=&M_{N}^2-\frac{\Delta^2}{4}\, .
\label{kin}
\ee
Here $\widetilde n^\mu$ and $n^\mu$ are arbitrary light-cone vectors
such that $\widetilde n\cdot n=1$ and
$\widetilde n \cdot \Delta_\perp=n \cdot \Delta_\perp=0$.
The momentum $k$ can be decomposed as follow:
\be
k^\mu=x\widetilde n^\mu +\beta n^\mu +k_\perp^\mu\, ,
\ee
where $\beta$ is fixed by
the
mass-shell conditions for
the partons. Let us note that we do not need here the explicit
expression for $\beta$ because it contributes to the DVCS amplitude
at the order $O(\Delta_\perp^2/Q^2)$. We are interested here
in the order $O(\Delta_\perp/Q)$.

Computing scattering of the virtual photon on the mass-shell
parton and neglecting terms of the order $O(\Delta_\perp^2/Q^2)$
and $M^2/Q^2$ one gets the
following expression for the DVCS amplitude in the parton model:
\be
\nonumber
M^{\mu\nu}\propto \int_{-1}^1  dx \int \frac{d^2 k_\perp}{(2\pi)^2}
\int d^4 z \ \delta(n\cdot z) \frac{1}{(x-\xi)(x+\xi)}
\exp[i(k\cdot z)]\\
\label{ampl}
\Biggl\{
\Tr\biggl[
(\Dirac k+\frac12 \Dirac \Delta) \Gamma^{\mu\nu}
(\Dirac k-\frac12 \Dirac \Delta) \Dirac n
 \biggr]\
\langle p'|
\bar \psi(-\frac{z}{2})\Dirac n \psi(\frac{z}{2}) |p \rangle \\
\nonumber
-\Tr\biggl[
(\Dirac k+\frac12 \Dirac \Delta) \Gamma^{\mu\nu}
(\Dirac k-\frac12 \Dirac \Delta) \Dirac n \gamma_5
 \biggr]\
\langle p'|
\bar \psi(-\frac{z}{2})\Dirac n\gamma_5 \psi(\frac{z}{2}) |p \rangle
\Biggr\}+ O(\frac{\Delta_\perp^2}{Q^2})\, ,
\ee
where
\be
\Gamma^{\mu\nu}=
\frac{\gamma^\mu(\Dirac k-\Dirac \Delta/2+\Dirac
q)\gamma^\nu}{(k-\Delta/2+q)^2+i0} +
\frac{\gamma^\nu(\Dirac k+\Dirac \Delta/2-\Dirac
q)\gamma^\mu}{(k+\Delta/2-q)^2+i0} \, .
\ee
Obviously,
the amplitude (\ref{ampl}) is transverse, $i.e.$
$(q-\Delta)^\mu M^{\mu\nu}=q^\nu M^{\mu\nu}=0$.
In the limit  $Q^2\to \infty$ and $\Delta_\perp^2 \ll Q^2$
the expression (\ref{ampl}) can be rewritten as:
\be
\nonumber
M^{\mu\nu}&=\frac 12 &\int_{-1}^1  dx
\Biggl\{\alpha^+(x,\xi) \
\biggl[ (\widetilde n^\mu n^\nu+\widetilde n^\nu n^\mu-g^{\mu\nu})
-\frac{4\xi}{Q^2} \Delta_\perp^\nu \widetilde n^\mu
\biggr]\  n^\alpha F_\alpha\\
\nonumber
&+&i\alpha^-(x,\xi)\
\biggl[
\varepsilon^{\mu \nu}_\perp+\frac{4\xi}{Q^2}
\widetilde n^\mu \varepsilon^{\Delta \nu}_\perp
\biggr]\ n^\alpha F_\alpha^{(5)}\\
\label{main2}
&+&
\biggl[n^\nu+\frac{8\xi^2}{Q^2}\widetilde n^\nu
\biggr]\  \biggl(\alpha^+(x,\xi)\ \bigl(F^{\mu_\perp}+
\frac{4\xi}{Q^2} \widetilde n^\mu (\Delta_\perp\cdot
F)\bigr)\\
\nonumber
&+&i\alpha^-(x,\xi)\ \bigl(
\varepsilon^{\mu \rho}_\perp
\ F_{\rho}^{(5)} +
\frac{4\xi}{Q^2} \widetilde n^\mu
\varepsilon_\perp^{\Delta\rho}F_{\rho}^{(5)}\bigr)
 \biggl) \\
\nonumber
&+&
n^\mu \biggl(\alpha^+(x,\xi)\ F^{\nu_\perp}
-i\alpha^-(x,\xi)\
\varepsilon^{\nu \rho}_\perp
\ F_{\rho}^{(5)} \biggl)\Biggr\}
+ O\biggl(\frac{\Delta_\perp^2}{Q^2} \biggr)
\, .
\ee
Here $
\varepsilon_\perp^{\mu\nu}=\varepsilon^{\mu\nu\alpha\beta}\
n_\alpha \widetilde n_\beta $.
We also defined:
\be
\nonumber
\alpha^{\pm}(x,\xi)&=& \frac{1}{x-\xi+i0} \pm \frac{1}{x+\xi-i0}\, .
\ee
Skewed distributions $F_\mu$ and $F_\mu^{(5)}$ are defined in terms of
bilocal quark operators on the light-cone\footnote{
The path ordered gauge link is assumed here. Although
in the parton model we do not have interactions with
the gluon fields the corresponding P-exponential can be obtained using
eikonal approximation for the quark propagator in the external gluon
field.}:

\be
\nonumber
F_{\mu}&=& \int \frac{d\lambda}{2\pi} e^{i x
\lambda} \langle p'| \bar \psi(-\frac \lambda 2 n)
\gamma_\mu
\psi(\frac
\lambda 2 n)| p \rangle  \, ,\\
F_{\mu}^{(5)}&=& \int \frac{d\lambda}{2\pi} e^{i x
\lambda} \langle p'| \bar \psi(-\frac \lambda 2 n)
\gamma_\mu \gamma_5
\psi(\frac
\lambda 2 n)| p \rangle  \, .
\label{bas3}
\ee

Generically, the expression (\ref{ampl})
contains operators off the
light-cone, but all of them can be reduced to
the operators (\ref{bas3})
using obvious operator identities like:
\be \bar \psi(y) \biggl[ \gamma_\alpha \nabla_\beta-
\gamma_\beta \nabla_\alpha\biggr] \psi(x)=
i\varepsilon_{\alpha\beta\rho\sigma}\
\bar \psi(y)  \gamma_\rho \gamma_5 \nabla_\sigma
\psi(x)\, ,
\label{UD}
\ee
which are satisfied owing to
the QCD equation of motion $\nabla \hskip-8pt/\psi(x)$.

Note that two first terms in the  expression
(\ref{main2})
coincide exactly with
the improved DVCS amplitude used in \cite{GuichonVander}. This part of
the amplitude is characterized by leading-twist skewed parton
distributions $n\cdot F$ and $n\cdot F^{(5)}$, other contributions are
characterized by new additional skewed parton distributions
$F_{\mu_\perp}$ and $F_{\mu_\perp}^{(5)}$.
The latter functions  are suppressed in differential cross section
of the reaction $e+N\to e'+\gamma+N$ by
only one power of the hard scale $1/Q$ \cite{Diehl,BM},
therefore their estimates are
important for the analysis of DVCS observables.
In principle, considering azimithal angle and $Q$
dependencies of the various spin and charge
asymmetries one should be able to disentangle the contributions of the
additional SPD's from the leading twist ones \cite{Diehl}.
Note also that in certain spin and azimuthal asymmetries the new
functions enter at the same order in $1/Q$ as the leading twist one,
as examples of such quantities are $\sin (2\phi)$ term in the lepton
spin asymmetry, and $const$ and
$\cos(2 \phi)$ term in the lepton charge asymmetry.
The measurements of such observables would give us an information
on the size of new SPD's.

It is easy to check that the amplitude (\ref{main2})
explicitly satisfies
the transversality conditions
$(q-\Delta)^\mu M^{\mu\nu}=q^\nu M^{\mu\nu}=0$.
In the case of
the pion target our
expression (\ref{main2}) coincides with the result obtained by Anikin
{\it et al.} \cite{Anikin}.
For the nucleon target we can write for skewed parton
distributions $F^\mu$ and $F^{\mu (5)}$, for instance, the following
decomposition:

\be
\nonumber
F^\mu &=& \bar u(p')\biggl\{ \gamma^\mu\ H(x,\xi,\Delta^2) +
\frac{i\sigma^{\mu\nu}\Delta_\nu}{2 M_N}E(x,\xi,\Delta^2) \\
&+&
\frac{
\Delta_\perp^\mu}{M_N}\biggl[
G_1(x,\xi,\Delta^2) +M_N \Dirac n\ G_2(x,\xi,\Delta^2)\biggr]+
\gamma_\perp^\mu
G_3(x,\xi,\Delta^2) +\ldots
\biggr\}u(p)\, ,\\
\nonumber
F^{\mu (5)} &=& \bar u(p')\biggl\{ \gamma^\mu\gamma_5
\ \widetilde H(x,\xi,\Delta^2) +
\frac{ \Delta^\mu}{2 M_N}\gamma_5
\widetilde E(x,\xi,\Delta^2) \\
&+&
\frac{\Delta_\perp^\mu}{M_N}\ \gamma_5
\widetilde G_1(x,\xi,\Delta^2)  +
\gamma_\perp^\mu \gamma_5
\widetilde G_2(x,\xi,\Delta^2)+
\Delta_\perp^\mu\ \Dirac n \gamma_5
\widetilde G_3(x,\xi,\Delta^2)+\ldots
\biggr\}u(p)\, ,
\ee
where ellipses stand for terms
which do not contribute to the order $1/Q$.
The two first functions in above equations
coincide with
the distributions introduced
in Ref.~\cite{Ji}. The functions $G_{i}(x,\xi,\Delta^2)$
$\widetilde G_i(x,\xi,\Delta^2)$ are
additional functions parameterizing the nucleon matrix element of
quark light-cone operator.
Obviously, these new functions satisfy
the following sum rule:
\be
\int_{-1}^1dx\ \widetilde G_{i}(x,\xi,\Delta^2)=0\,,\quad
\int_{-1}^1dx\ G_{i}(x,\xi,\Delta^2)=0\, .
\ee
The additional function $G_1(x,\xi,\Delta^2)$ receive
a contribution from  the so-called D-term in the
double distributions parametrization of light-ray operators
\cite{PW}.
This contribution to the function
$G_1(x,\xi,\Delta^2)$ has the form:

\be G_1(x,\xi,\Delta^2)=\frac{1}{2\xi}\ D\biggl(\frac x\xi,
\Delta^2\biggr)\, .  \ee

In the forward limit the SPD's $F_\mu$ and $F_\mu^{(5)}$ are:
\be
\nonumber
F_\mu &\to& 2\bigl\{ f_1(x) \widetilde n_\mu + M_N^2 f_4(x)
n_\mu \bigr\}\\
\nonumber
F_\mu^{(5)} &\to& 2\bigl\{
g_1(x) \widetilde n_\mu\ (n\cdot S) + g_T(x) S_{\perp\mu} +
g_3(x) M_N^2 n_\mu\ (n\cdot S)
\bigr\}\, .
\ee
Hence, we observe that in the forward limit the SPD's
$F_\mu$ and $F_\mu^{(5)}$ are reduced to the combinations of twist-2
parton distributions and higher twist distributions $f_4(x), g_3(x)$
and $g_T^{tw3}(x)$.
The functions $F_\mu$ and $F_\mu^{(5)}$ can be computed in the chiral
quark-soliton model using methods of Refs.~\cite{cqsm}.

Very interesting sum rules can be derived if we consider the second
Mellin moments of the distributions $G_i$ and $\widetilde G_i$.
With help of identities like (\ref{UD}) one can derive
relations between the second Mellin moments. Here we restrict ourselves
to the following relations:

\be
\label{g3}
\int^1_{-1} dx\ x\ G_3(x,\xi)&=&-\frac 12
\int^1_{-1} dx\ x\ \biggl[H(x,\xi)+E(x,\xi)\biggr]+
\frac 12 \int^1_{-1} dx\ \widetilde H(x,\xi)\, ,\\
\label{g2}
\int^1_{-1} dx\ x\
\widetilde G_2(x,\xi)&=&-\frac 12
\int^1_{-1} dx\ x\ \widetilde H(x,\xi)+
\frac{\xi^2}{2} \int^1_{-1} dx\  \biggl[H(x,\xi)+E(x,\xi)\biggr]\, .
\ee
The first relation in the forward limit can be related to the
quark orbital momentum because:
\be
\lim_{\Delta^\mu \to 0}\int_{-1}^1 dx\ x\ G_3(x,\xi)=
-J_q+ \frac 12
\Delta q =- L_q\, .
\label{newsr}
\ee
Here $L_q$ is the quark orbital momentum contribution to the
proton spin. In derivation we used   Ji's sum rule \cite{Ji}.
We see that the distribution $G_3$ is deeply related to the spin
structure of the nucleon.

The second relation (\ref{g2}) is the non-forward
generalization of the Efremov-Leader-Teryaev sum rule \cite{ELT},
it reduces to the ELT sum rule in the forward limit, because in the
forward limit the SPD's $\widetilde H$ and $\widetilde G_2$
are reduced to the spin distributions
$g_1$ and $ g_2$ correspondingly.

As the final remark we note that
the new functions $G_i$ and $\widetilde G_i$
can be related to the leading twist functions $H,\widetilde H$
and $E, \widetilde E$ if one neglects the contributions of
operators of type $\bar \psi G \psi$. The method of derivation is
analogous to derivation of the Wandzura--Wilczek type of
relations between twist-2 and twist-3
rho-meson distributions amplitudes \cite{BB}. Recently such
relations were derived in ref.~\cite{BM}.

\section*{\normalsize {\bf Discussion}}

We have demonstrated
that the handbag contribution to the DVCS amplitude up to the order
$1/Q$ depends not only on the leading twist skewed parton distributions
$(n\cdot F)$  and $(n\cdot F^{(5)})$ but also on
the additional functions $F_{\rho_\perp}$ and $F^{(5)}_{\rho_\perp}$.
These additional contributions are
suppressed by only one power of the hard
scale--$1/Q$ in the differential cross section. The estimates of these
contributions are important for the extraction of the
leading twist skewed parton
densities $H,\widetilde H$ and $E,\widetilde E$ from observables
\cite{Belitsky} and measuring the Ji's sum rule \cite{Ji}.
Also it is encouraging that the new functions can be also related
to the spin structure of the nucleon, see eqs.~(\ref{g3},\ref{g2}).
This can give additional possibility to extract quark orbital momentum
from DVCS observables.

Note also that in our analysis we used the approximation
$\Delta_\perp^2\ll Q^2$, natural for the parton model.
An account of the QCD evolution may lead effectively to a breakdown of
this approximation and further complicate description of DVCS.
Indeed, in the QCD ladder virtualities of the partons in the rungs of the
ladder close to the nucleon are much smaller than $Q^2$. They are rather
close to $Q_0^2$. Hence the accuracy of neglecting $\Delta_\perp^2$
in these propagators is $\sim \Delta_\perp^2/Q_0^2$, not
$\sim \Delta_\perp^2/Q^2$.
The effects due to evolution of distributions $F_{\mu_\perp}$
and $F_{\mu_\perp}^{(5)}$ will be studied elsewhere.

\section*{\normalsize {\bf Acknowledgements}}
We are grateful to V.Yu.~Petrov  for sharing
with us his ideas and valuable contributions to calculations.
Discussions with L.~Frankfurt, K.~Goeke, N.~Kivel,
D.~M\"uller, P.V.~Pobylitsa,
A.~Radyushkin, A.~Sch\"afer, J.~Soffer, O.~Teryaev,
M.~Vanderhaeghen are greatly acknowledged.
M.V.P. is thankful to C.~Weiss for many
interesting discussions.
Critical remarks of M.~Diehl and N.~Kivel were of big
help for us.

\end{document}